\documentclass[12pt]{article}
\usepackage{amsmath}
\usepackage{graphicx}
\title{Jordan blocks and Gamow-Jordan eigenfunctions associated to a 
double pole of the $S-$matrix}
\author{E. Hern\'andez, A. J\'auregui$^{\dagger}$ and  A. 
Mondrag\'on\thanks{This 
work was partially supported by CONACyT M\'exico 
under contract number 32238-E and by DGAPA-UNAM contract No. PAPIIT: 
IN125298}\\
Instituto de F\'{\i}sica, UNAM,
  Apdo. Postal 20-364, \\ 01000 M\'exico D.F., \ M\'exico
\and $^{\dagger}$Departamento de F\'{\i}sica, Universidad de Sonora,
\\ Apdo. Postal 
1626,  Hermosillo, Sonora, M\'exico }
\date{}

\begin{document}

\maketitle
\begin{abstract}
  An accidental degeneracy of resonances gives rise to a double pole
  in the scattering matrix, a double zero in the Jost function and a
  Jordan chain of length two of generalized Gamow-Jordan
  eigenfunctions of the radial Schr\"odinger equation. The generalized
  Gamow-Jordan eigenfunctions are basis elements of an expansion in
  bound and resonant energy eigenfunctions plus a continuum of
  scattering wave functions of complex wave number. In this
  biorthonormal basis, any operator $f(H^{(\ell)}_{r})$ which is a
  regular function of the Hamiltonian is represented by a complex
  matrix wich is diagonal except for a Jordan block of rank two. The
  occurrence of a double pole in the Green's function, as well as the
  non-exponential time evolution of the Gamow-Jordan generalized
  eigenfunctions are associated to the Jordan block in the complex
  energy representation.
  
  Keywords: Non-relativistic scattering theory; Multiple resonances;
  Resonance reactions; Berry's phase.
  
  PACS: 03.65.Nk; 33.40.+f; 24.30.-v; 03.65.Bz

\end{abstract}

\section{Introduction}

Lately, the interference effects of resonances, the crossing and
anticrossing properties of the energies and widths of two unbound
levels and the occurrence of a double pole of the scattering matrix
have aroused a great deal of interest. Some interesting examples of
interfering unbound two level systems are the $T =1, T=0, J^{\pi} =
2^{+}$ doublet in $^{8}$Be\cite{one,two,three}, the $T=1, T=0$ doublet
of $\rho$ and $\omega$ messons and the $\sigma-K_{s}$ doublet of
neutral sigma and $K$ mesons\cite{two,four,five,six,seven}.  A variety
of widely differing systems where double poles can occur have been
identified, such as autoionizing states in complex atoms\cite{eight}
and atomic states in intense laser fields\cite{nine,ten}. The problem
of the degeneracy of resonances also arises naturally in connection
with the Berry phase of resonant
states\cite{eleven,twelve,thirteen,fourteen} which was recently
measured by the Darmstadt group\cite{fifteen}. Some examples of simple
quantum mechanical systems with double poles in the scattering matrix
have been recently described. Vanroose et al.,\cite{sixteen} examined
the formation of complex double poles of the $S$-matrix in a two
channel model with square well potentials. Recently, Hern\'andez et
al.,\cite{seventeen} investigated a one channel model with two
spherical concentric cavities bounded by $\delta-$function barriers
and showed that a double pole of the $S-$matrix can be induced by
tuning the parameters of the model; Vanroose generalized this model to
the case of two finite width barriers\cite{eighteen}. The formal
theory of multiple pole resonances and resonant states in the rigged
Hilbert space formulation of quantum mechanics was developed by Bohm
et al.,\cite{nineteen} and by Antoniou et al.,\cite{twenty}.

In the present paper, we deal with the problem of multiple poles of
the scattering matrix and the generalized complex energy
eigenfunctions associated with them in the framework of the theory of
the analytic properties of the radial wave functions.

The plan of this paper is as follows. In sections 2 and 3, we
introduce some basic concepts and fix the notation by way of a short
reminder of resonances and resonant states in the theory of the
analytic properties of the radial wave functions. Sections 4 and 5 are
devoted to a short discussion of the no-crossing rule for bound states
and its non applicability to resonant states. In section 6, we show
that a double pole of the scattering wave function (double zero of the
Jost function) is associated to a chain of length two of Gamow-Jordan
generalized eigenfunctions and derive explicit expressions for this
generalized eigenfunctions in terms of the outgoing wave Jost
solution, the Jost function and its derivatives evaluated at the
double pole. We also show that the Gamow-Jordan generalized
eigenfunctions in the Jordan chain are elements of a complete set of
states containing the real (bound states) and complex (resonant state)
energy eigenfunctions plus a continuum of scattering wave functions of
complex wave number.  In section 7 we derive expansion theorems
(spectral representations) for operators $f(H^{(\ell)}_{r})$ which are
regular functions of the radial Hamiltonian $H^{(\ell)}_{r}$ and show
that, in this basis, the operator $f(H^{(\ell)}_{r})$ is represented
by a complex matrix which is diagonal except for a Jordan block of
rank two associated to the double zero of the Jost function and the
corresponding Jordan chain of generalized Gamow-Jordan
eigenfunction.We give the normalization and orthogonality rules for
the generalized eigenfunctions in the Jordan chain associated to the
double pole of the Green's function in section 8. We end our paper
with a summary of results and some conclusions in section 9.

\section{Regular and physical solutions of the radial equation}

The non-relativistic scattering of a spinless particle by a short
ranged potential $v(r)$ is described by the solution of a
Schr\"odinger equation. When the potential is rotationally invariant,
the wave function is expanded in partial waves and one is left with
the radial equation

\begin{equation}\label{uno}
\frac{d^{2}\phi_{\ell }\left( k,r\right) }{dr^{2}}+\left[ k^{2}-
\frac{\ell\left( \ell +1\right) }{r^{2}}-v\left( r\right) \right] 
\phi_{\ell }\left(k,r\right) = 0. 
\end{equation}
As is usually done when discussing the analytic properties of the
solutions of (1) as functions of $k$, rather than starting by defining
the physical solutions $\psi^{(+)}_{\ell}(k,r)$, we define the regular
and irregular solutions of (\ref{uno}) by boundary conditions which
lead to simple properties as functions of $k$. The regular solution
$\phi_{\ell}(k,r)$ is uniquely defined by the boundary condition
\cite{twentyone}

\begin{eqnarray}\label{dos}
\lim_{r\rightarrow 0}(2\ell+1)!! r^{-\ell-1}\phi_{\ell}(k,r) = 1,
\end{eqnarray}
\noindent
$\phi_{\ell}(k,r)$ may be expressed as a linear combination of two
independent, irregular solutions of (\ref{uno}) which behave as
outgoing and incoming waves at infinity,

\begin{eqnarray}\label{tres}
\phi_{\ell}(k,r) = \frac{1}{2}ik^{-\ell
  -1}\left[f_{\ell}(-k)f_{\ell}(k,r) -
  (-1)^{\ell}f_{\ell}(k)f_{\ell}(-k,r)\right],
\end{eqnarray}
where $f_{\ell}(-k,r)$ is an outgoing wave at infinity defined by the
boundary condition
\begin{eqnarray}\label{cuatro}
\lim_{r\rightarrow \infty}\exp(-ikr)f_{\ell}(-k,r) = (+i)^{\ell}
\end{eqnarray}
and $f_{\ell}(k,r)$ is an incoming wave at infinity related to
$f_{\ell}(-k,r)$ by

\begin{eqnarray}\label{cinco}
f_{\ell}(k,r) = (-1)^{\ell}f^{*}_{\ell}(-k,r)
\end{eqnarray}
for $k$ real and non-vanishing.

The Jost function $f_{\ell}(-k)= f_{\ell}(-k,0)$ is given by

\begin{eqnarray}\label{seis}
f_{\ell}(-k) = (-1)^{\ell}k^{\ell}W[f_{\ell}(-k,r),\phi_{\ell}(k,r)]
\end{eqnarray}
where $W[f,g] = fg'-f'g$ is the Wronskian. The Jost function
$f_{\ell}(-k)$, has zeroes (roots) on the imaginary axis and in the
lower half of the complex $k$ plane.

When the first and second absolute moments of the potential exist, and
the potential decreases at infinity faster than any exponential (e.g.
if $v(r)$ has a gaussian tail or if it vanishes identically beyond a
finity radius) the functions $f_{\ell}(-k)$, $\phi_{\ell}(k,r)$, and
$k^{\ell}f_{\ell}(-k,r)$, for fixed $r>0$, are entire function of
$k$\cite{twentyone}.

Therefore, the derivatives of these functions with respect to the wave
number $k$ exist and are entire functions of $k$ for all finite values
of $k$ in the complex $k$-plane.

The differential equations satisfied by the derivatives of the
functions $\phi_{\ell}(k,r)$ and $f_{\ell}(-k,r)$ with respect to $k$
are obtained from (\ref{uno}) taking derivatives with respect to $k$
in both sides of the equation,

\begin{eqnarray}\label{siete}
\frac{d^{2}\dot{\phi}_{\ell}(k,r)}{dr^{2}} + \Bigl[k^{2} -
\frac{\ell(\ell+1)}{r^{2}} - v(r)\Bigr]\dot{\phi}_{\ell}(k,r) = 
-2k\phi_{\ell}(k,r),
\end{eqnarray}

\begin{eqnarray}\label{ocho}
\frac{d^{2}\ddot{\phi}_{\ell}(k,r)}{dr^{2}} + \Bigl[k^{2} -
\frac{\ell(\ell+1)}{r^{2}} - v(r)\Bigr]\ddot{\phi}_{\ell}(k,r) = 
-4k\dot{\phi}_{\ell}(k,r) -2\phi_{\ell}(k,r),
\end{eqnarray}
in (\ref{siete}) and (\ref{ocho}) we have used the notation
$\dot{\phi}_{\ell}(k,r) = d\phi_{\ell}(k,r)/dk$. Similar expressions
are valid for the derivatives with respect to $k$ of the outgoing wave
solutions $f_{\ell}(-k,r)$.

The scattering wave function $\psi^{(+)}_{\ell}(k,r)$ is the solution
of equation (\ref{uno}) which vanishes at the origin and behaves at
infinity as the sum of a free incoming spherical wave of unit incoming
flux plus a free outgoing spherical wave,

\begin{eqnarray}\label{nueve}
\psi^{(+)}_{\ell}(k,0) = 0
\end{eqnarray}
and
 
\begin{eqnarray}\label{diez}
\lim_{r\rightarrow \infty }\left\{ \psi^{(+)} _{\ell }( k, r)
  -\left[\hat{h}^{\left( -\right) }_{\ell}( k, r) - S_{\ell}(k) 
\hat{h}^{\left( +\right) }_{\ell}\left( k,r\right) \right] 
\right\} =0. 
\end{eqnarray}
In this expression $\hat{h}^{(-)}_{\ell}(k,r)$ and
$\hat{h}^{(+)}_{\ell}(k,r)$ are Ricatti-Hankel functions that describe
incoming and outgoing waves respectively, $S_{\ell}(k)$ is the
scattering matrix.

Hence, the scattering wave function $\psi^{(+)}_{\ell}(k,r)$ and the
regular solution are related by

\begin{eqnarray}\label{once}
\psi^{(+)}_{\ell}(k,r) =
\frac{k^{\ell+1}\phi_{\ell}(k,r)}{f_{\ell}(-k)},
\end{eqnarray}
and the scattering matrix is given by

\begin{eqnarray}\label{doce}
S_{\ell}(k) = \frac{f_{\ell}(k)}{f_{\ell}(-k)}.
\end{eqnarray}

The complete Green's function for outgoing particles or resolvent of
the radial equation may also be written in terms of the regular
solution $\phi_{\ell}(k,r)$ and the irregular solution
$f_{\ell}(-k,r)$ which behaves as an outgoing wave at infinity

\begin{eqnarray}\label{trece}
G^{(+)}_{\ell}(k;r,r') =
(-1)^{\ell+1}k^{\ell}\frac{\phi_{\ell}(k,r_{<})f_{\ell}
(-k,r_{>})}{f_{\ell}(-k)}.
\end{eqnarray}

\section{Bound and resonant state eigenfunctions}

Bound and resonant state energy eigenfunctions are the solutions of
(\ref{uno}) which vanish at the origin

\begin{eqnarray}\label{catorce}
u_{n\ell}(k_{n},0) = 0,
\end{eqnarray}
and at infinity satisfy the boundary condition

\begin{eqnarray}\label{quince}
\lim_{r\rightarrow\infty}\left[\frac{1}{u_{n\ell}(k_{n},r)}
\frac{du_{n\ell}(k_{n},r)}{dr}
  - ik_{n}\right] = 0,
\end{eqnarray}
where $k_{n}$ is a zero of the Jost function,

\begin{eqnarray}\label{diesiseis}
f_{\ell}(-k_{n}) = 0.
\end{eqnarray}
From equations (\ref{uno}) and (\ref{tres}) we verify that all roots
(zeroes) of the Jost function are associated to energy eigenfunctions
of the Schr\"odinger equation.

Bound state eigenfunctions are associated to the zeroes of
$f_{\ell}(-k)$ which lay on the positive imaginary axis $k^{2}_{s}
=-\kappa^{2}_{s} < 0$, while resonant or Gamow state eigenfunctions
are associated to the zeroes of the Jost function which lay in the
fourth quadrant of the complex $k$-plane.

From (\ref{tres}), (\ref{cuatro}) and (\ref{diesiseis}), bound states
and Gamow or resonance eigenfunctions are related to the regular
solution $\phi_{\ell}(k,r)$ by

\begin{eqnarray}\label{diesisiete}
u_{n\ell}(k_{n},r) = N_{n\ell}^{-1}\phi_{\ell}(k_{n},r)
\end{eqnarray}
where $N_{n\ell}$ is a normalization constant. Due to the vanishing of
$f_{\ell}(-k_{n})$, $\phi_{\ell}(k_{n},r)$ is now proportional to the
outgoing wave solution, $f_{\ell}(-k_{n},r)$, of (\ref{uno}). Hence,

\begin{eqnarray}\label{diesiocho}
u_{n\ell}(k_{n},r) = N^{-1}_{n\ell}\frac{i}{2}\frac{(-1)^{\ell
    +1}}{k^{\ell+1}}f_{\ell}(k_{n})f_{\ell}(-k_{n},r).
\end{eqnarray}

This expression shows, in a very explicit way, that the Gamow state
eigenfunctions $u_{n\ell}(k_{n},r)$ with $k_{n} = \kappa_{n}
-i\gamma_{n}$ and $\kappa_{n} > \gamma_{n} > 0$, are solutions of
(\ref{uno}) which vanish at the origin and asymptotically behave as
purely outgoing waves which oscillate between envelopes that increase
exponentially with $r$, the corresponding energy eigenvalues
${\cal{E}}_{n}$ are complex with Re ${\cal{E}}_{n}>$
Im${\cal{E}}_{n}$.

The bound state eigenfunctions $u_{s\ell}(k_{s},r)$ are also solutions
of (\ref{uno}) which satisfy the boundary conditions (\ref{catorce})
and (\ref{quince}), but, in this case, $k_{s} = i\kappa_{s}$ with,
$\kappa_{s} > 0 $, which means that asymptotically the outgoing wave
of imaginary argument, $f_{\ell}(-k_{s},r)$, decreases exponentially
with $r$ and the energy eigenvalue ${\cal{E}}_{s}$ is real and
negative.

\section{The no-crossing rule for bound states.}

In the case of bound states, the normalization constant is related to
the derivative of the Jost function evaluated at $k_{s}$ and it may
also be expressed as a normalization integral. The zero of the Jost
function is on the positive imaginary axis, and the bound state
eigenfunction is cuadratically integrable (for time reversal invariant
forces $\phi_{\ell}(i\kappa_{s},r)$ is real). R.G. Newton gives the
following expression\cite{twentyone}

\begin{eqnarray}\label{diesinueve}
N^{2}_{s\ell} =
\frac{1}{i4k_{s}^{2(\ell+1)}}\left(\frac{df_{\ell}(-k)}{dk}
\right)_{k_{s}}f_{\ell}(k_{s})
  = \int^{\infty}_{0}|\phi_{\ell}(k_{s},r)|^{2}dr.
\end{eqnarray}

Since the normalization integral is positive and the function
$f_{\ell}(k)$ is regular at $k_{s} = i\kappa_{s}$, the derivative of
the Jost function evaluated at $k_{s} = i\kappa_{s}$ cannot vanish.
Therefore, the zero of $f_{\ell}(-k)$ at $k_{s} = i\kappa_{s}$ must be
simple. The corresponding pole in $G^{(+)}_{\ell}(k;r,r')$,
$\psi^{(+)}_{\ell}(k,r)$ and $S_{\ell}(k)$ must also be simple.

It follows that, in the absence of symmetry, the real, negative energy
eigenvalues of the radial equation for a one channel problem cannot be
degenerate.

\section{Crossing of resonant states}

In the case of a resonant state, the zero of the Jost function
$f_{\ell}(-k)$ lies in the fourth quadrant of the complex $k$-plane,

\begin{eqnarray}\label{veinte}
k_{n} = \kappa_{n} - i\gamma_{n},
\end{eqnarray}
with $\kappa_{n} > \gamma_{n} > 0$.

The resonant or Gamow eigenfunction $\phi_{\ell}(k_{n},r)$ is an
outgoing spherical wave of complex wave number $k_{n}$ and angular
momentum $\ell$. Therefore, for large values of $r$,
$\phi_{\ell}(k_{n},r)$ oscillates between envelopes that grow
exponentially with $r$. Hence, the integrals over $r$ must be properly
defined. This may be done by means of a gaussian regulator and a
limiting procedure\cite{twentytwo} T.
Berggren\cite{twentythree,twentyfour} gives the following expression

\begin{eqnarray}\label{veintiuno}
\frac{1}{i4k_{n}^{2(\ell+1)}}\left(\frac{df_{\ell}(-k)}{dk}
\right)_{k_n}f_{\ell}(k_{n}) = \lim_{\nu\rightarrow 0}
\int^{\infty}_{0}\exp(-\nu r^{2})\phi^{2}_{\ell}(k_{n},r) dr
\end{eqnarray}
The integral in the right hand side is a complex number and it may
vanish.

Since $f_{\ell}(k_{n})$ has no zeroes in the lower half of the complex
$k$-plane, the left hand side of equation (\ref{veintiuno}) vanishes
only when $(df_{\ell}(-k)/dk)_{k_{n}}$ vanishes. Then, we have two
possibilities,

i) When $(df_{\ell}(-k)/dk)_{k_{n}}$ does not vanish, $f(-k)$, has a
simple zero at $k=k_{n}$, the integral in the right hand side of
equation (\ref{veintiuno}) does not vanish and the normalization
constant, $N^{2}_{n\ell}$, occurring in (\ref{diesisiete}) is given by
(\ref{veintiuno}).

ii) When 
\begin{eqnarray}\label{veintidos}
\left(\frac{df_{\ell}(-k)}{dk}\right)_{k_{n}} = 0,
\end{eqnarray}
the integral in the right hand side of (\ref{veintiuno}) vanishes,

\begin{eqnarray}\label{veintitres}
\lim_{\nu\rightarrow 0}\int^{\infty}_{0}\exp(-\nu
r^{2})\phi^{2}_{\ell}(k_{n},r) dr = 0
\end{eqnarray}
and the Jost function $f_{\ell}(-k)$ has a multiple zero at $k=k_{n}$.
In this case, the Green's function $G^{(+)}_{\ell}(k;r,r')$, the
scattering wave function $\psi^{(+)}_{\ell}(k,r)$ and the scattering
matrix $S_{\ell}(k)$ have a multiple pole at $k = k_{n}$. The
normalization constant of the Gamow eigenfunction is no longer given
by (\ref{veintiuno}).

Furthermore, it will be shown below that when $f_{\ell}(-k)$ has a
multiple zero ( a multiple resonant pole of rank $r$ in
$G^{(+)}_{\ell}(k;r,r')$, $\psi^{(+)}_{\ell}(k,r,)$ and $S_{\ell}(k)$)
the corresponding complex energy eigenvalues are degenerate even in
the absence of symmetry. That is, the no-crossing rule does not hold
for resonant eigenstates.

\section{Completeness and the expansion in complex resonance energy
  eigenfunctions}

In this section, it will be shown that associated to a double zero of
the Jost function (double pole of the scattering wave function
$\psi^{(+)}_{\ell}(k,r)$, the Green's function
$G^{(+)}_{\ell}(k;r,r')$ and the scattering matrix $S_{\ell}(k)$) there
is a chain of generalized Gamow-Jordan eigenfunctions which together
with the bound state and resonant state eigenfunctions form a
biorthonormal set which may be completed with a continuum of
scattering wave functions of complex wave number.

Given two square integrable and very well behaved functions $\Phi (r)$
and $\chi (r)$ which decrease at infinity faster than any exponential,
the completeness of the orthonormal set of bound state and scattering
solutions of the radial Schr\"odinger equation\cite{twentyone} allows
us to write

\begin{eqnarray}\label{veinticuatro}
<\Phi|\chi> &=& \sum_{s \ bound
  \ states}<\Phi|v_{s,\ell}><v_{s,\ell}|\chi>  \cr
 &+& \frac{2}{\pi}\int^{\infty}_{0}<\Phi|\psi^{(+)}_{\ell}(k')>
<\psi^{(+)}_{\ell}(k')|\chi>dk'
\end{eqnarray}
where $<\Phi|\chi>$ is the standard Dirac bracket notation

\begin{eqnarray}\label{veinticinco}
<\Phi|\chi> = \int^{\infty}_{0}\Phi^{*}(r)\chi(r)dr.
\end{eqnarray}

We shall assume that the Jost function $f_{\ell}(-k)$ has a double
zero at $k=k_{m}$ in the fourth quadrant of the complex $k'-$plane,
all other zeroes of $f_{\ell}(-k')$ in that quadrant being simple.
Then, the scattering function $\psi^{(+)}_{\ell}(k',r)$ as function of
$k'-$complex, has one double resonance pole at $k'=k_{m}$ and simple
resonance poles at $k=k_{n}$, $n=1,2 .......  m-1, m+1 ...,$ all in
the fourth quadrant of the complex $k'-$plane. The function
$\psi^{(+)*}(k',r)$ is regular and has no poles in the lower half of
the $k'-$plane.

\begin{figure}
\begin{center}
\includegraphics[width=290pt,height=260pt]{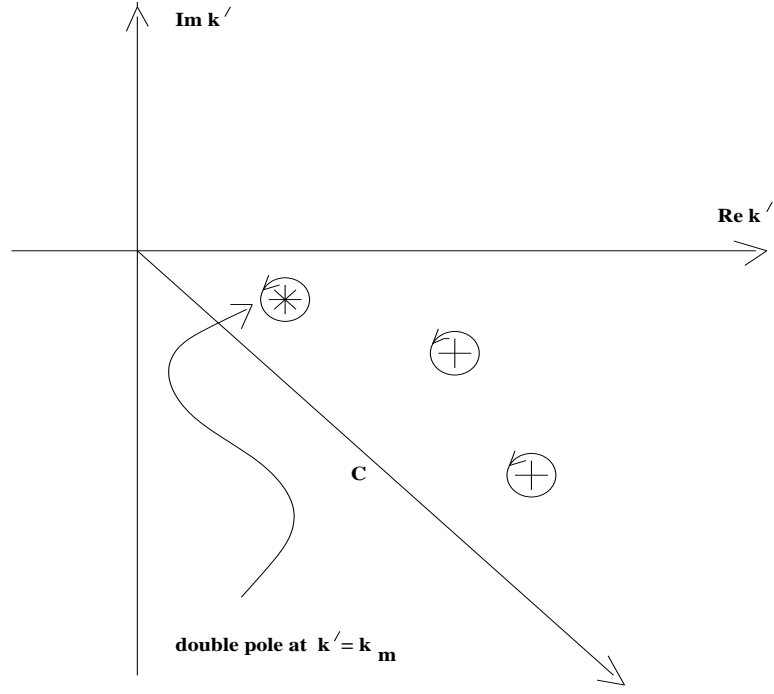}
\caption{Integration contour ${\bf C}$ in the complex $k' -$plane.}
\end{center}
\end{figure}
In order to make explicit the contribution of the resonant states to
the expansion in eigenfunctions, the integration contour in the second
term in the right hand side of (\ref{veinticuatro}) is deformed as
shown in Fig. 1.

When the deformed contour $C$ crosses over resonant poles, the 
theorem of the residue gives

\begin{eqnarray}\label{veintiseis}
<\Phi|\chi> &=& \sum_{s \ bound
  \ states}<\Phi|v_{s,\ell}><v_{s,\ell}|\chi> \cr
&+& \sum_{all   \ resonance \ poles}2\pi i
  Res\Bigl[\frac{2}{\pi}<\Phi|\psi^{(+)}_{\ell}(k')>
<\psi^{(+)}_{\ell}(k')|\chi>\Bigr] \nonumber \\
  &+&  \frac{2}{\pi}\int_{C}<\Phi|\psi^{(+)}_{\ell}(k')>
<\psi^{(+)}_{\ell}(k')|\chi>dk'.
\end{eqnarray}

The residues may be readily computed from equations (\ref{once}) and
(\ref{veintiseis}).

When $f_{\ell}(-k')$ has a simple zero at $k'=k_{n}$,

\begin{eqnarray}\label{veintisiete}
2\pi i
Res\Bigl[\frac{2}{\pi}<\Phi|\psi^{(+)}_{\ell}(k')>
<\psi^{(+)}_{\ell}(k')|\chi>\Bigr]_{k'= k_{n}}
= \nonumber \\ 4i Res
\Bigl[\frac{<\Phi|\phi_{\ell}(k')><\phi_{\ell}(k')|\chi>k^{'2(\ell +
    1)}}{(k'-k_{n})\Bigl(\frac{df_{\ell}(-k')}{dk'}
\Bigr)_{k_{n}}f_{\ell}(k')}\Bigr]_{k'= k_{n}} = \nonumber \\
\frac{1}{\frac{f_{\ell}(k_{n})}{4i k^{2(\ell +1)}_{n}}
\Bigl(\frac{df_{\ell}(-k')}{dk'}\Bigr)_{k_{n}} }
\Bigl[<\Phi|\phi_{\ell}(k')>\Bigr]_{k'=k_{n}}\Bigl[<\phi_{\ell}(k')|
\chi>\Bigr]_{k'= k_{n}}
\end{eqnarray}
where
\begin{eqnarray}\label{veintiocho}
\Bigl[<\Phi|\phi_{\ell}(k')>\Bigr]_{k'= k_{n}} = \lim_{k'\rightarrow
  k_{n}}\int^{\infty}_{0}\Phi^{*}(r)\phi_{\ell}(k',r)dr,
\end{eqnarray}
and, 
\begin{eqnarray}\label{veintinueve}
\Bigl[<\phi_{\ell}(k')|\chi>\Bigr]_{k'= k_{n}} = \lim_{k'\rightarrow
  k_{n}}\int^{\infty}_{0}\phi_{\ell}(k',r)\chi(r)dr.
\end{eqnarray}
since $\phi_{\ell}(k',r)$ is real and bounded for $k'$ real, the
integrals in (\ref{veintiocho}) and (\ref{veintinueve}) exist.

Furthermore, since $\phi_{\ell}(k_{n},r)$ is an outgoing wave which
oscillates between envelopes that grow exponentially at infinity and
$\Phi(r)$ and $\chi(r)$ are very well behaved functions of $r$ that
decrease at infinity faster than any exponential, the integrals of the
products $\Phi^{*}(r)\phi_{\ell}(k_{n},r)$ and
$\phi_{\ell}(k_{n},r)\chi(r)$ also exist, and we may take the limit
indicated in the right hand side of equations (\ref{veintiocho}) and
(\ref{veintinueve}) under the integration sign.

Therefore,

\begin{eqnarray}\label{treinta}
2\pi i
Res\Bigr[\frac{2}{\pi}<\Phi|\psi^{(+)}_{\ell}(k')><\psi^{(+)}_{\ell}
(k')|\chi>\Bigr]_{k'= k_{n}}
= <\Phi|u_{n\ell}(k_{n})><u_{n\ell}(k_{n})|\chi>
\end{eqnarray}
where the notation means,

\begin{eqnarray}\label{treintaiuno}
<\Phi|u_{n\ell}(k_{n})> = \int^{\infty}_{0}\Phi^{*}(r)
u_{n\ell}(k_{n},r)dr
\end{eqnarray}
and

\begin{eqnarray}\label{treintaidos}
<u_{n\ell}(k_{n})|\chi> = \int^{\infty}_{0}u_{n\ell}(k_{n},r)
\chi(r)dr,
\end{eqnarray}

The Gamow eigenfunction or normal mode, $u_{n\ell}(k_{n},r)$, is given
by (\ref{diesiocho}) and the normalization constant $N_{n\ell}$ is
given by

\begin{eqnarray}\label{treintaitres}
N^{2}_{n\ell} =
\frac{1}{i4k_{n}^{2(\ell+1)}}f_{\ell}(k_{n})
\bigl(\frac{df_{\ell}(-k')}{dk'}\bigr)_{k_{n}},
\end{eqnarray}
in agreement with Berggren's result given in equation
(\ref{veintiuno}).

When the Jost function  $f_{\ell}(-k')$ has a double zero at
$k'=k_{m}$,  $\psi^{(+)}_{\ell}(k',r)$ has a double pole at $k' =
k_{m}$, 

\begin{eqnarray}\label{treintaicuatro}
\psi^{(+)}_{\ell}(k',r) = \frac{\phi_{\ell}(k',r)k^{'(\ell + 1)}}
{(k' -  k_{m})^{2}g_{\ell m}(k')}.
\end{eqnarray}
the function $g_{\ell m}(k')$ is regular at $k'=k_{m}$ and may be
expanded as

\begin{eqnarray}\label{treintaicinco}
g_{\ell m}(k') =
\frac{1}{2}\bigl(\frac{d^{2}f_{\ell}(-k')}{dk^{'2}}\bigr)_{k_{m}} +
\frac{1}{6}(k'-k_{m})\bigl(\frac{d^{3}f_{\ell}(-k')}{dk^{'3}}
\bigr)_{k_{m}} + ....
\end{eqnarray}
with

\begin{eqnarray}\label{treintaiseis}
\left(\frac{d^{2}f_{\ell}(-k')}{dk^{'2}}\right)_{k_{m}} \neq 0.
\end{eqnarray}
The function $\psi^{(+)*}_{\ell}(k',r')$ is regular at $k'=k_{m}$,
since $f_{\ell}(k')$ has no zeroes in the lower half of the complex
$k'-$plane,

\begin{eqnarray}\label{treintaisiete}
\psi^{(+)*}_{\ell}(k',r') =
\frac{\phi_{\ell}(k',r')k^{'(\ell+1)}}{f_{\ell}(k')}.
\end{eqnarray}
Thus, the residue of the term $(2/\pi)
<\Phi|\psi^{(+)}_{\ell}(k')><\psi^{(+)}_{\ell}(k')|\chi>$ at the
double pole in $k'=k_{m}$ is obtained from the Cauchy integral 
formula as

\begin{eqnarray}\label{treintaiocho}
2\pi i
Res\Bigl[\frac{2}{\pi}<\Phi|\psi^{(+)}_{\ell}(k')><
\psi^{(+)}_{\ell}(k')|\chi>\Bigr]_{k'=k_{m}} = \nonumber \\
4i Res\Bigl[\frac{<\Phi|\phi_{\ell}(k')><\phi_{\ell}(k')|
\chi>k^{'2(\ell+1)}}{(k'-k_{m})^{2}g_{\ell
    m}(k')f_{\ell}(k')}\Bigr]_{k'= k_{m}} =  \nonumber \\
4i\Big[\frac{d}{dk'}\Big(\frac{<\Phi|\phi_{\ell}(k')>
<\phi_{\ell}(k')|\chi>k^{'2(\ell+1)}}{g_{\ell  m}(k')
f_{\ell}(k')} \Big) \Big]_{k' = k_{m}}.
\end{eqnarray}

After computing the derivative indicated in (\ref{treintaiocho}) and
rearranging some terms, we obtain

\begin{eqnarray}\label{treintainueve}
2\pi i
Res\Big[\frac{2}{\pi}<\Phi|\psi^{(+)}_{\ell}(k')><\psi^{(+)}_{\ell}
(k')|\chi>\big]_{k' = k_{m}} = \nonumber \\
\frac{1}{{\cal N}_{m\ell}^{2}}
\Bigl[<\Phi|\hat{\phi}_{\ell}(k_{m})><\phi_{\ell}(k_{m})|\chi> +
<\Phi|\phi_{\ell}(k_{m})><\hat{\phi}_{\ell}(k_{m})|\chi>\Bigr],
\end{eqnarray}
where, according to (\ref{diesisiete}), $\phi_{\ell}(k_{m},r)$ is the
non-normalized Gamow eigenfunction, and $\hat{\phi}_{\ell}(k_{m},r)$
is a generalized Gamow-Jordan eigenfunction or abnormal mode given by

\begin{eqnarray}\label{cuarenta}
\hat{\phi}_{\ell}(k_{m},r) = \frac{d\phi_{\ell}(k_{m},r)}{d{\cal
    E}_{m}} + C_{\ell}(k_{m})\phi_{\ell}(k_{m},r),
\end{eqnarray}

\noindent
${\cal E}_{m}$ is the complex energy eigenvalue, ${\cal E}_{m} =
(\hbar^{2}/2\mu) k_{m}^{2}$, and the constant factor
$C_{\ell}(k_{m})$, multiplying $\phi_{\ell}(k_{m},r)$ in
equation (\ref{cuarenta}), is

\begin{eqnarray}\label{cuarentaiuno}
C_{\ell}(k_{m}) = \frac{2\mu}{\hbar^{2}}\frac{1}{2k_{m}}\Bigl[
\frac{\ell+1}{k_{m}}-\frac{1}{2}\frac{1}{f_{\ell}(k_{m})}
\frac{df_{\ell}(k_{m})}{dk_{m}}-\frac{1}{6}\Bigl(\frac{d^{2}f_{\ell}
(-k')}{dk^{'2}}\Bigr)^{-1}_{k_{m}}\Bigl(\frac{d^{3}f_{\ell}(-k')}
{dk^{'3}}\Bigr)_{k_{m}}\Bigr]
\end{eqnarray}

The normalization constant ${\cal N}^{2}_{ml}$ is now 

\begin{eqnarray}\label{cuarentaidos}
{\cal N}_{m\ell}^{2} =\Bigl(\frac{2\mu}{\hbar^{2}}\Bigr)
\frac{1}{16ik_{m}^{2\ell+3}}f_{\ell}(k_{m})\Bigl(\frac{d^{2}f_{\ell}
(-k')}{dk^{'2}}\Bigr)_{k_{m}}.
\end{eqnarray}

The expression (\ref{treintainueve}) suggests the following
normalization rule for the chain of Gamow-Jordan generalized
eigenfunctions belonging to a double zero of the Jost function

\begin{eqnarray}\label{cuarentaitres}
u_{m\ell}(k_{m},r) = \frac{1}{{\cal N}_{m\ell}}\phi_{\ell}(k_{m},r),
\end{eqnarray}
and

\begin{eqnarray}\label{cuarentaicuatro}
\hat{u}_{m\ell}(k_{m},r) = \frac{1}{{\cal
    N}_{m\ell}}\hat{\phi}_{\ell}(k_{m},r).
\end{eqnarray}
Substitution of (\ref{cuarentaitres}) and (\ref{cuarentaicuatro}) in
(\ref{treintainueve}) gives

\begin{eqnarray}\label{cuarentaicinco}
2\pi i Res
\Bigl[\frac{2}{\pi}<\Phi|\psi^{(+)}_{\ell}(k')><\psi^{(+)}_{\ell}
(k')|\chi>\Bigr]_{k'= k_{m}}  = \nonumber \\
<\Phi|\hat{u}_{m\ell}(k_{m})><u_{m\ell}(k_{m})|\chi> +
<\Phi|u_{m\ell}(k_{m})><\hat{u}_{m\ell}(k_{m})|\chi>
\end{eqnarray}
where, the notation means,

\begin{eqnarray}\label{cuarentaiseis}
<\Phi|\hat{u}_{m\ell}(k_{m})> =
\int^{\infty}_{0}\Phi^{*}(r)\hat{u}_{m\ell}(k_{m},r)dr
\end{eqnarray}
and

\begin{eqnarray}\label{cuarentaisiete}
<\hat{u}_{m\ell}(k_{m})|\chi> = \int^{\infty}_{0}\hat{u}_{m,\ell}
(k_{m},r)\chi(r) dr,
\end{eqnarray}
$\hat{u}_{m\ell}(k_{m},r)$ is defined in (\ref{cuarentaicuatro}).

Finally, substitution of the expressions (\ref{treinta}) and
(\ref{cuarentaicinco}) in (\ref{veintiseis}) gives the following
expansion

\begin{eqnarray}\label{cuarentaiocho}
<\Phi|\chi> &=& \sum_{s \ bound \
  states}<\Phi|v_{s\ell}><v_{s\ell}|\chi> 
+ \sum_{n\neq  m \
  resonances} <\Phi|u_{n\ell}><u_{n\ell}|\chi> \nonumber \\
&+& <\Phi|\hat{u}_{m\ell}(k_{m})><u_{m\ell}(k_{m})|\chi> 
+  <\Phi|u_{m\ell}(k_{m})><\hat{u}_{m\ell}(k_{m})|\chi> \nonumber \\
&+& \frac{2}{\pi}\int_{c}<\Phi|\psi^{(+)}_{\ell}(k')><\psi^{(+)}_{\ell}
(k')|\chi>dk'.
\end{eqnarray}
This expression shows that, when the Jost function has many simple
zeroes and one double zero in the fourth quadrant of the complex
k-plane, the Gamow eigenfunctions $u_{n\ell}(k_{m},r)$ associated to
simple zeroes of the Jost function and the chain
$\{u_{m\ell}(k_{m},r), \hat{u}_{m\ell}(k_{m},r)\}$ of Gamow-Jordan
generalized eigenfunctions\cite{twentyfive,twentysix,twentyseven}
associated to the double pole of the Jost function are basis elements
of an expansion in generalized bound and resonant state eigenfunctions
plus a continuum of scattering functions of complex wave values $k'$.

Omitting the arbitrary function $\Phi(r)$ in (\ref{cuarentaiocho}), we
obtain the complex basis expansion of an arbitrary square integrable
and well behaved function $\chi(r)$

\begin{eqnarray}\label{cuarentainueve}
\chi(r) &=& \sum_{s \ bound \ states}v_{s\ell}(r)<v_{s\ell}|\chi>
+ \sum_{n\neq m}u_{n\ell}(k_{n},r)<u_{n\ell}|\chi> \nonumber \\
&+& \hat{u}_{m\ell}(k_{m},r)<u_{m\ell}|\chi> +
u_{m\ell}(k_{m},r)<\hat{u}_{m\ell}|\chi> \nonumber \\ 
&+& \frac{2}{\pi}\int_{c}\psi^{(+)}_{\ell}(k',r)<\psi^{(+)}_{\ell}
(k')|\chi>dk'.
\end{eqnarray}

In this expression $u_{n\ell}(k_{n},r)$ are the Gamow eigenfunctions
representing decaying states associated to simple resonance poles of
the scattering wave function $\psi^{(+)}_{\ell}(k,r)$, the matrix
$S(k)$ and the Green's function $G^{(+)}(k;r,r')$. The set
$\{u_{m\ell}(k_{m},r), \hat{u}_{m\ell}(k_{m},r)\}$ is a Jordan chain
of lenght two of generalized Gamow-Jordan eigenfunctions associated to
the double pole of the scattering matrix $S(k)$ and the Green's
function $G^{(+)}(k;r,r')$ at $k=k_{m}$. The last term in the right
hand side of (\ref{cuarentaiocho}) and (\ref{cuarentainueve}) is the
background integral defined along the integration contour shown in 
Fig 1.

\section{Jordan blocks in the complex energy basis}

Once it has been established that the Gamow eigenfunctions
$u_{n\ell}(k_{n},r)$ and the Jordan chain $\{u_{m\ell}(k_{m},r),
\hat{u}_{m\ell}(k_{m},r)\}$ of generalized Gamow-Jordan eigenfunctions
are elements of the basis set of eigenfunctions in the expansions
(\ref{cuarentaiocho}) and (\ref{cuarentainueve}), we may represent any
operator $f(H^{(\ell)}_{r})$, which is a regular function of the
Hamiltonian $H^{(\ell)}_{r}$, in terms of its matrix elements in this
basis.

Let us start by deriving an expression for the action of
$f(H^{(\ell)}_{r})$ on the generalized Gamow-Jordan eigenfunction
$\hat{u}_{m\ell}(k_{m},r)$. With this purpose in mind, let us write
the eigenvalue equation satisfied by $u_{m\ell}(k_{m},r)$ as

\begin{eqnarray}\label{cincuenta}
H^{\ell}_{r}u_{m\ell}(k_{m},r) = {\cal{E}}_{m}u_{m\ell}(k_{m},r),
\end{eqnarray}
where,

\begin{eqnarray}\label{cincuentaiuno}
H^{\ell}_{r} = -\frac{\hbar^{2}}{2\mu}\Bigl[\frac{d^{2}}{dr^{2}} - 
v(r) - \frac{\ell(\ell+1)}{r^{2}}\Bigr],
\end{eqnarray}
$v(r)$ is a well behaved short ranged potential which satisfies the
conditions stated in section 1. Now, let us consider a holomorphic
function $f({\cal{E}})$ of the complex variable ${\cal{E}}$, such
that,

\begin{eqnarray}\label{cincuentaidos}
f({\cal{E}}) = \sum_{j=0}^{\infty}a_{j}{\cal{E}}^{j},
\end{eqnarray}
the coefficients $a_{j}$ are independent of ${\cal{E}}$.

Then, from (\ref{cincuenta}) and (\ref{cincuentaidos}),

\begin{eqnarray}\label{cincuentaitres}
f(H^{(\ell)}_{r})u_{m\ell}(k_{m},r) = f({\cal{E}}_{m})
u_{m\ell}(k_{m},r).
\end{eqnarray}

Taking derivatives with respect to the eigenvalue ${\cal{E}}_{m}$ in
both sides of (\ref{cincuentaitres}), we obtain,

\begin{eqnarray}\label{cincuentaicuatro}
f(H_{r}^{(\ell)})\frac{\partial u_{m\ell}(k_{m},r)}{\partial 
{\cal{E}}_{m}} = f({\cal{E}}_{m})\frac{\partial
  u_{m\ell}(k_{m},r)}{\partial{\cal{E}}_{m}} + \frac{\partial
  f({\cal{E}}_{m})}{\partial{\cal{E}}_{m}}u_{m\ell}(k_{m},r).
\end{eqnarray}
From this equation and the definition, equations (\ref{cuarenta}),
(\ref{cuarentaiuno}) and (\ref{cuarentaicuatro}), of
$\hat{u}_{m\ell}(k_{m},r)$, it follows inmediatly that

\begin{eqnarray}\label{cincuentaicinco}
f(H^{(\ell)}_{r})\hat{u}_{m\ell}(k_{m},r) =
f({\cal{E}}_{m})\hat{u}_{m\ell}(k_{m},r) + \frac{\partial
  f({\cal{E}}_{m})}{\partial{\cal{E}}_{m}}u_{m}(k_{m},r).
\end{eqnarray}
Notice that a necessary and sufficient condition for the existence of
$\partial u_{m\ell}(k_{m},r)/\partial{\cal{E}}_{m}$ is the vanishing
of $\Bigl(df(-k)/dk\Bigr)_{k_{m}}$.

The rule stated in equation (\ref{cincuentaicinco}) permits us to
calculate the action of $f(H^{(\ell)}_{r})$ on the generalized
Gamow-Jordan vectors occurring in the complex basis expansions
(\ref{cuarentaiocho}) and (\ref{cuarentainueve}).

Now, we can write the operator $f(H^{(\ell)}_{r})$ in terms of its
matrix elements in the complex energy basis. This may be done by
acting with $f(H^{(\ell)}_{r})$ on the left in both sides of equation
(\ref{cuarentainueve}),

\begin{eqnarray}\label{cincuentaiseis}
f(H^{(\ell)}_{r})\chi(r) &=&
\sum_{s}f({\cal{E}}_{s})v_{s\ell}(r)<v_{s\ell}|\chi> +
\sum_{n\neq  m}f({\cal{E}}_{n})u_{n\ell}(k_{n},r)<u_{n\ell}|\chi> 
\nonumber \\
&+&\Bigl(f({\cal{E}}_{m})\hat{u}_{m\ell}(k_{m},r) +
\frac{\partial f({\cal{E}}_{m})}{\partial {\cal{E}}_{m}}u_{m\ell}
(k_{m},r)\Bigr)<u_{m\ell}|\chi> \nonumber \\
&+& f({\cal{E}}_{m})u_{m\ell}(k_{m},r)<\hat{u}_{m\ell}|\chi> 
\nonumber \\
&+& \frac{2}{\pi}\int_{c}f({\cal{E}}')\phi^{(+)}_{\ell}(k',r)
<\phi^{(+)}_{\ell}(k')|\chi>dk'. 
\end{eqnarray}
Multiplying both sides of (\ref{cincuentaiseis}) by $\Phi^{*}(r)$ and
integrating over $r$, we get,

\begin{eqnarray}\label{cincuentaisiete}
<\Phi|f(H^{(\ell)}_{r})|\chi> &=&
\sum_{s}<\Phi|v_{s\ell}>f({\cal{E}}_{s})<v_{s\ell}|\chi> +
\sum_{n \neq m}<\Phi|u_{n\ell}>f({\cal{E}}_{n})<u_{n\ell}|\chi>
\nonumber \\
&+& <\Phi|\hat{u}_{m\ell}>f({\cal{E}}_{m})<u_{m\ell}|\chi> 
+ <\Phi|u_{m\ell}>\Bigl(f({\cal{E}}_{m}) <\hat{u}_{m\ell}|\chi>
\nonumber \\
&+&
\frac{\partial f({\cal{E}}_{m})}{\partial {\cal{E}}_{m}}<u_{m\ell}
|\chi>\Bigr) 
+ \frac{2}{\pi}\int_{c}<\Phi|\psi^{(+)}_{\ell}(k')>f({\cal{E}}')
<\psi^{(+)}_{\ell}(k')|\chi>dk'. \nonumber  \\
\end{eqnarray}

To simplify the notation, suppose that the system has no bound states
only resonances and that the first two resonances are degenerate.
Rearranging equation (\ref{cincuentaisiete}) in matrix form, we get

\begin{eqnarray}\label{cincuentaiocho}
<\Phi|f(H^{(\ell)}_{r})|\chi> = 
\Bigl(<\Phi|u_{1\ell}>,<\Phi|\hat{u}_{1\ell}>,<\Phi|u_{3\ell}>,
.....\Bigr) \times  \nonumber  \\   
\begin{pmatrix}
f({\cal{E}}_{1}) & \frac{\partial f({\cal{E}}_{1})}{\partial
  {\cal{E}}_{1}} & 0 & 0 & 0 & 0 & . & . \\
0 & f({\cal{E}}_{1}) & 0 & 0 & 0 & 0 & . & . \\
0 & 0 & f({\cal{E}}_{3}) & 0 & 0 & 0 & . & . \\
0 & 0 & 0 & f({\cal{E}}_{4}) & 0 & 0 & . & .\\
. & . & . & . & . & . & . & . \\
. & . & . & . & . & . & . & . 
\end{pmatrix}
\begin{pmatrix}
<\hat{u}_{1\ell}|\chi > \\
<u_{1\ell}|\chi> \\
<u_{3\ell}|\chi> \\
. \\
. \\
.
\end{pmatrix}
 \nonumber  \\
+ \frac{2}{\pi}\int_{c}<\Phi|\psi^{(+)}_{\ell}(k')>f({\cal{E}}')
<\psi^{(+)}_{\ell}(k')|\chi> dk'.
\end{eqnarray}

In this matrix representation of $f(H^{(\ell)}_{r})$\footnote{From the
  way it was derived, it is evident that the matrix in equation
  (\ref{cincuentaiocho}) represents the action of $f(H^{(\ell)}_{r})$
  as an operator on the space of continuous antilinear functionals on
  the Schwarz space of very well behaved test functions.}, the upper
left $2\times 2$ submatrix is a Jordan block of rank
two\cite{twentyfive,twentysix,twentyseven} associated to the chain of
Gamow-Jordan generalized eigenfunctions $\{\hat{u}_{1\ell}(k_{1},r),
u_{1\ell}(k_{1},r)\}$ belonging to the double zero of the Jost
function $f_{\ell}(-k)$ (double pole of the scattering matrix and the
Green's function). Except for this $2\times 2$ block, this matrix is
diagonal with the eigenvalues $f({\cal{E}}_{n})$ in the diagonal
entries. Simple zeroes of the Jost function correspond to simple
(non-repeated) eigenvalues of $f(H^{(\ell)}_{r})$ while the double
zeroe of $f_{\ell}(-k)$ correspond to the twice repeated (degenerate)
eigenvalue $f({\cal{E}}_{1})$ occuring in the Jordan block. The
off-diagonal non-vanishing element in this block is $\partial
f({\cal{E}}_{1})/\partial {\cal{E}}_{1}$.

The difference in physical dimensions of the off-diagonal and the
diagonal entries in the $2\times 2$ Jordan block is compensated by the
difference in normalization of the Gamow-Jordan chain
$\{\hat{u}_{1\ell}(k_{1},r), u_{1\ell}(k_{1}, r)\}$ and the Gamow
eigenfunctions $u_{n\ell}(k_{n},r)$ $(n=3,4,....)$ which are
normalized according to (\ref{cuarentaitres}, \ref{cuarentaicuatro})
and (\ref{diesisiete}, \ref{diesiocho}, \ref{treintaitres})
respectively.

It will be instructive to consider some simple examples.  

We first choose $f(H^{(\ell)}_{r}) = H^{(\ell)}_{r}$. Then, from
(\ref{cincuentaiocho}) we obtain,

\begin{eqnarray}\label{cincuentainueve}
<\Phi|H^{(\ell)}_{r}|\chi> = 
\Bigl(<\Phi|u_{1\ell}>,<\Phi|\hat{u}_{1\ell}>,<\Phi|u_{3\ell}>,
.....\Bigr) \times  \nonumber  \\   
\begin{pmatrix}
{\cal{E}}_{1} & 1 & 0 & 0 & 0 & 0 & . & . \\
0 & {\cal{E}}_{1} & 0 & 0 & 0 & 0 & . & . \\
0 & 0 & {\cal{E}}_{3} & 0 & 0 & 0 & . & . \\
0 & 0 & 0 & {\cal{E}}_{4} & 0 & 0 & . & .\\
. & . & . & . & . & . & . & . \\
. & . & . & . & . & . & . & . 
\end{pmatrix}
\begin{pmatrix}
<\hat{u}_{1\ell}|\chi > \\
<u_{1\ell}|\chi> \\
<u_{3\ell}|\chi> \\
. \\
. \\
.
\end{pmatrix}
 \nonumber  \\
+ \frac{2}{\pi}\int_{c}<\Phi|\psi^{(+)}_{\ell}(k')>
{\cal{E}}'<\psi^{(+)}_{\ell}(k')|\chi> dk'.
\end{eqnarray}

From this example, it is evident that in a degeneracy of two
resonances in the absence of symmetry, the degenerate complex
eigenvalue ${\cal{E}}_{1}$ occurs twice in the spectral representation
of the radial Hamiltonian $H^{(\ell)}_{r}$ given in
(\ref{cincuentainueve}), while there is only one Gamow eigenvector or
normal mode, $u_{1\ell}(k_{1},r)$, associated to the degeneracy. This
is so, because the Gamow-Jordan generalized eigenfunction or abnormal
mode, $\hat{u}_{1\ell}(k_{1},r)$, is not an eigenfunction of the
radial Hamiltonian $H^{(\ell)}_{r}$. This is a generic property of
this kind of degeneracy which may be stated in slightly more formal
terms as follows: In a degeneracy of resonances in the absence of
symmetry, the algebraic multiplicity is always larger than the
geometric multiplicity.  Here, we mean by algebraic multiplicity of a
degeneracy, $\mu_{a}$, the number of times the degenerate complex
eigenvalue is repeated, and, by geometric multiplicity of the
degeneracy, $\mu_{g}$, the dimensionality of the subspace spanned by
the eigenvectors associated to the degenerate
eigenvalue\cite{twentyfive,twentysix,twentyseven}.

Then,

\begin{eqnarray}\label{sesenta}
\mu_{a} > \mu_{g}.
\end{eqnarray}

Let us consider now, the complex energy representation of the
resolvent operator. In this case $f(H^{(\ell)}_{r}) =
1/(E-H^{(\ell)}_{r})$. Then, from (\ref{cincuentaiocho}), we obtain,

\begin{eqnarray}\label{sesentaiuno}
<\Phi|\frac{1}{E-H^{(\ell)}_{r}}|\chi> = 
\Bigl(<\Phi|u_{1\ell}>,<\Phi|\hat{u}_{1\ell}>,<\Phi|u_{3\ell}>,
.....\Bigr) \times  \nonumber  \\   
\begin{pmatrix}
\frac{1}{E-{\cal{E}}_{1}} & \frac{1}{(E-{\cal{E}}_{1})^{2}}   & 
0 & 0 & 0 & 0 & . & . \\
0 & \frac{1}{E-{\cal{E}}_{1}}  & 0 & 0 & 0 & 0 & . & . \\
0 & 0 & \frac{1}{E-{\cal{E}}_{3}} & 0 & 0 & 0 & . & . \\
0 & 0 & 0 & \frac{1}{E-{\cal{E}}_{4}} & 0 & 0 & . & .\\
. & . & . & . & . & . & . & . \\
. & . & . & . & . & . & . & . 
\end{pmatrix}
\begin{pmatrix}
<\hat{u}_{1\ell}|\chi > \\
<u_{1\ell}|\chi> \\
<u_{3\ell}|\chi> \\
. \\
. \\
.
\end{pmatrix}
 \nonumber  \\
+ \frac{2}{\pi}\int_{c}<\Phi|\psi^{(+)}_{\ell}(k')>
\frac{1}{(E-{\cal{E}}^{'})}<\psi^{(+)}_{\ell}(k')|\chi> dk'.
\end{eqnarray}

It may easily be verified that, when we delete the arbitrary functions
$\Phi(r)$ and $\chi(r)$ in this expression, the resulting expansion
for $<r|\frac{1}{E-H^{(\ell)}_{r}}|r'>$ is just the expansion in
resonance eigenfunctions of the complete Green's function

\begin{eqnarray}\label{sesentaidos}
G^{(+)}_{\ell}(k;r,r') &=& \frac{\hbar^{2}}{2\mu}\Bigl[\sum_{s \ 
bound \ states}\frac{v_{s\ell}(k,r)v^{*}_{s\ell}(k,r')}
{E + |E_{s}|} \cr 
 &+& \sum_{n \neq m\ resonant \
  states}\frac{u_{n\ell}(k_{n},r)u_{n\ell}(k_{n},r')}{E -
  {\cal E}_{n}} \cr 
 &+& \frac{u_{m\ell}(k_{m},r)u_{m\ell}(k_{m},r')}
{(E - {\cal E}_{m})^{2}}   \cr
 &+& \frac{u_{m\ell}(k_{m},r)\hat{u}_{m\ell}(k_{m},r') +
  \hat{u}_{m\ell}(k_{m},r)u_{m\ell}(k_{m},r')}{(E - {\cal
  E}_{m})}\Bigr] \cr
 &+& \frac{2}{\pi}\int_{C}\frac{\psi^{(+)}_{\ell}(k',r)
\psi^{(+)*}_{\ell}(k',r')}{(k^{2}
  - k^{'2})} dk'.
\end{eqnarray}

The occurrence of the double pole in $G^{(+)}_{\ell}(k;r,r')$, as
function of the complex energy, is thus associated to the occurrence
of a Jordan block of rank two in the complex basis representation of
the resolvent operator and a Jordan chain of Gamow-Jordan generalized
eigenfunctions $\{\hat{u}_{1\ell}(k_{1},r), u_{1\ell}(k_{1},r)\}$
associated to the double zero of the Jost function.

Finally, let us consider the time evolution operator $\exp{(-iHt)}$.
For each fixed value of the angular momentum, it will be enough to
consider the operator $f(H^{(\ell)}_{r}) = \exp{(-iH^{(\ell)}_{r}t)}$.
In this case, from equation (\ref{cincuentaiocho})

\begin{eqnarray}\label{sesentaitres}
<\Phi|\exp{(-iH^{(\ell)}_{r}t)}|\chi> = 
\Bigl(<\Phi|u_{1\ell}>,<\Phi|\hat{u}_{1\ell}>,<\Phi|u_{3\ell}>,
.....\Bigr) \times \nonumber  \\   
\begin{pmatrix}
\exp{(-i{\cal{E}}_{1}t)} & -it \exp{(-i{\cal{E}}_{1}t)}   & 
0 & 0  & . & . \\
0 & \exp{(-i{\cal{E}}_{1}t)}  & 0 & 0 & . & . \\
0 & 0 & \exp{(-i{\cal{E}}_{3}t)} & 0 & . & . \\
0 & 0 & 0 & \exp{(-i{\cal{E}}_{4}t)} &  . & .\\
. & . & . & . & . & .   \\
. & . & . & . & . & .   
\end{pmatrix}
\begin{pmatrix}
<\hat{u}_{1\ell}|\chi > \\
<u_{1\ell}|\chi> \\
<u_{3\ell}|\chi> \\
. \\
. \\
.
\end{pmatrix}
\nonumber \\
+ \frac{2}{\pi}\int_{c}<\Phi|\psi^{(+)}_{\ell}(k')>
\exp{(-i{\cal{E}}'t)}<\psi^{(+)}_{\ell}(k')|\chi> dk'.
\end{eqnarray}

As in the previous examples, the time evolution operator is non
diagonal in the complex energy basis representation. The time
evolution of the Jordan chain of Gamow-Jordan generalized
eigenfunctions $\{\hat{u}_{1\ell}(k_{1},r), u_{1\ell}(k_{1},r)\}$ is
given by a Jordan block of $2\times 2$ with an exponential time
dependence in the diagonal entries and a first order polynomial times
an exponential in the off-diagonal entry. Hence, the time evolution of
the Gamow-Jordan generalized eigenfunction or abnormal mode is a
superposition of the abnormal mode $\hat{u}_{1\ell}(k_{1},r)$ evolving
exponentially in time plus the normal mode $u_{1\ell}(k,r)$ evolving
according to the product of a first order polynomial times an
exponential time evolution factor. The time evolution of the normal
mode $u_{1\ell}(k_{1},r)$ in the Gamow-Jordan chain
$\{\hat{u}_{1\ell}(k_{1},r), u_{1\ell}(k_{1},r)\}$, as well as the
time evolution of all other normal modes $u_{n\ell}(k_{n},r)$
associated to the simple zeroes of the Jost function (simple poles of
the scattering matrix) is purely exponential.

\section{Orthogonality and normalization integrals for Gamow-Jordan
eigenfunction.}

As in the case of bound and resonant state eigenfunctions associated
with simple poles of the Green's function, we may derive orthogonality
and normalization rules for the Gamow-Jordan eigenstates in terms of
regularized integrals of the generalized Gamow-Jordan eigenfunctions.
Following the same procedure as in
Berggren\cite{twentythree,twentyfour}, it may be shown that, when
$f_{\ell}(-k')$ has a double zero at $k'=k_{m}$, the following
relations are valid,

\begin{eqnarray}\label{sesentaicuatro}
\hspace{-0.6cm}\frac{1}{i8k_{m}^{2(\ell +
    1)}}f_{\ell}(k_{m})\Bigl(\frac{d^{2}f_{\ell}(-k')}
{dk^{'2}}\Bigr)_{k'=k_{m}}
    = \lim_{\nu\rightarrow 0}\int^{\infty}_{0}e^{-\nu
    r^{2}}\frac{d\phi_{\ell}(k_{m},r)}{dk_{m}}\phi_{\ell}(k_{m},r) dr
\end{eqnarray}
and

\begin{eqnarray}\label{sesentaicinco}
\frac{1}{i8k_{m}^{2(\ell+1)}}f_{\ell}(k_{m})\Bigl[\frac{1}{3}
\Bigl(\frac{d^{3}f_{\ell}(-k')}{dk^{'3}}\Bigr)_{k'= k_{m}} - 
\Bigl(\frac{d^{2}f_{\ell}(-k')}{dk^{'2}}\Bigr)_{k'= k_{m}}\Bigl
(\frac{2(\ell + 1)}{k_{m}} - \frac{1}{f_{\ell}(k_{m})}
\frac{df_{\ell}(k_{m})}{dk_{m}}\Bigr)\Bigr]
 \nonumber  \\ 
= \lim_{\nu\rightarrow 0}\int^{\infty}_{0}e^{-\nu
  r^{2}}\Bigl(\frac{d\phi_{\ell}(k_{m},r)}{dk_{m}}\Bigr)^{2} dr.
\end{eqnarray}

From the expression (\ref{cuarentaiuno}) for $C_{\ell}(k_{m})$ and
equations (\ref{sesentaicuatro}) and (\ref{sesentaicinco}), it 
follows that,

\begin{eqnarray}\label{sesentaiseis}
\lim_{\nu\rightarrow 0}\int^{\infty}_{0}e^{-\nu
  r^{2}}\Bigl(\frac{d\phi_{\ell}(k_{m},r)}{dk_{m}}\Bigr)^{2} dr +
  \nonumber \\
2C_{\ell}(k_{m})\frac{\hbar^{2}k_{m}}{\mu}\Bigl(
\lim_{\nu\rightarrow 0}\int^{\infty}_{0}e^{-\nu
    r^{2}}\frac{d\phi_{\ell}(k_{m},r)}{dk_{m}}\phi_{\ell}(k_{m},r)
  dr\Bigr) = 0,
\end{eqnarray}
which may be rewritten as,

\begin{eqnarray}\label{sesentaisiete}
\lim_{\nu\rightarrow 0}\int^{\infty}_{0}e^{-\nu
    r^{2}}\Bigl[\frac{d\phi_{\ell}(k_{m},r)}{d{\cal E}_{m}} + 
C_{\ell}(k_{m})\phi_{\ell}(k_{m},r)\Bigr]^{2}dr = 
C^{2}_{\ell}(k_{m})\lim_{\nu\rightarrow 0}\int^{\infty}_{0}e^{-\nu
  r^{2}}\phi^{2}_{\ell}(k_{m},r) dr,
\end{eqnarray}
but, according to equation (\ref{veintidos}) and (\ref{veintitres}),
when $f_{\ell}(-k)$ has a double zero at $k=k_{m}$, the integral in
the right hand side of (\ref{sesentaisiete}) vanishes. Therefore, the
integrand in the left hand side of (\ref{sesentaisiete}) is the square
of the generalized Jordan-Gamow eigenfunction and the relation
(\ref{sesentaicinco}) translates into

\begin{eqnarray}\label{sesentaiocho}
\lim_{\nu\rightarrow 0}\int^{\infty}_{0}e^{-\nu
  r^{2}}\hat{\phi}^{2}_{\ell}(k_{m},r) dr = 0
\end{eqnarray}
which shows that also the regularized integral of the square of the
generalized Gamow-Jordan eigenfunction vanishes.

An expression for the normalization constant ${\cal N}^{2}_{m\ell}$ in
terms of a normalization integral may be obtained from
(\ref{sesentaicuatro}),

\begin{eqnarray}\label{sesentainueve}
{\cal N}^{2}_{m\ell} = \lim_{\nu\rightarrow 0}\int^{\infty}_{0}
e^{-\nu r^{2}}\frac{d\phi_{\ell}(k_{m},r)}{d{\cal E}_{m}}
\phi_{\ell}(k_{m},r) dr,
\end{eqnarray}
writing $d\phi_{\ell}/d{\cal E}_{m}$ in terms of
$\hat{\phi}_{\ell}(k_{m},r)$ and recalling that the integral of
$\phi^{2}_{\ell}(k_{m},r)$ vanishes, we get,

\begin{eqnarray}\label{setenta}
{\cal N}^{2}_{m\ell} = \lim_{\nu\rightarrow 0}\int^{\infty}_{0}e^{-\nu
  r^{2}}\hat{\phi}_{\ell}(k_{m},r)\phi_{\ell}(k_{m},r) dr,
\end{eqnarray}
which shows that the right hand side of (\ref{setenta}) is the
normalization integral for the Gamow-Jordan generalized eigenfunctions
associated with a double pole degeneracy of resonances with ${\cal
  N}^{2}_{m\ell}$ as given in (\ref{cuarentaidos}). However, it is
convenient to note that this expression does not fix the normalization
rule for $\phi_{\ell}(k_{m},r)$ and $\hat{\phi}_{\ell}(k_{m},r)$ in a
unique way. Since $\phi_{\ell}(k_{m},r)$ and
$\hat{\phi}_{\ell}(k_{m},r)$ are linearly independent, they have
different dimensions and its product has no obvious interpretation in
terms of observable quantities, therefore, there is no a priori reason
to normalize both functions with the same normalization constant.
Thus, we still have the freedom to write (\ref{setenta}) as

\begin{eqnarray}\label{setentaiuno}
\lim_{\nu \rightarrow 0}\int^{\infty}_{0}e^{-\nu
  r^{2}}\Bigl(\frac{X_{m}}{{\cal
  N}_{m\ell}}\hat{\phi}_{\ell}(k_{m},r)\Bigr)\Bigl(\frac{1}{X_{m}{\cal
  N}_{m\ell}}\phi_{\ell}(k_{m},r)\Bigr)dr = 1,
\end{eqnarray}
where ${\cal N}^{2}_{m\ell}$ is given in (\ref{cuarentaidos}) and
$X_{m}$ is a non-vanishing real or complex number that we associate
with the double pole singularity of $G^{(+)}_{\ell}(k;r,r')$ at
$k=k_{m}$. Therefore, a more general normalization rule for the Gamow
and Gamow-Jordan generalized eigenfunction that the one proposed in
(\ref{cuarentaidos}), (\ref{cuarentaitres}) and
(\ref{cuarentaicuatro}) would be

\begin{eqnarray}\label{setentaidos}
u_{m\ell}(k_{m},r) = \frac{1}{X_{m}{\cal
    N}_{m\ell}}\phi_{\ell}(k_{m},r)
\end{eqnarray}
and
\begin{eqnarray}\label{setentaitres}
\hat{u}_{m\ell}(k_{m},r) = \frac{X_{m}}{{\cal
    N}_{m\ell}}\hat{\phi}_{\ell}(k_{m},r).
\end{eqnarray}
With this normalization, the orthogonality and normalization integrals
for the generalized Gamow-Jordan eigenfunction associated to a double
pole of the Green's function, equations (\ref{veintitres}),
(\ref{sesentaiocho}) and (\ref{setenta}), take the form

\begin{eqnarray}\label{setentaicuatro}
\lim_{\nu\rightarrow 0}\int^{\infty}_{0}e^{-\nu
  r^{2}}u^{2}_{m\ell}(k_{m},r)dr = 0
\end{eqnarray}

\begin{eqnarray}\label{setentaicinco}
\lim_{\nu\rightarrow 0}\int^{\infty}_{0}e^{-\nu
  r^{2}}\hat{u}^{2}_{m\ell}(k_{m},r)dr = 0
\end{eqnarray}
and

\begin{eqnarray}\label{setentaiseis}
\lim_{\nu\rightarrow 0}\int^{\infty}_{0}e^{-\nu
  r^{2}}u_{m\ell}(k_{m},r)\hat{u}_{m\ell}(k,r)dr = 1.
\end{eqnarray}

The form of these orthogonality and normalization conditions is
independent of the value of the constat $X_{m}$. However, if the
Gamow-Jordan generalized eigenfunction are normalized according to
(\ref{setentaidos}) and (\ref{setentaitres}) the expression for the
residue at the double pole of $G^{(+)}_{\ell}(k;r,r')$ would be
explicitly dependent on $X_{m}$, since a factor $X^{2}_{m}$ will
appear multiplying the term $u_{m\ell}(k_{m},r)u_{m\ell}(k_{m},r')$ 
in the expression for the residue at the double pole of
$G^{(+)}_{\ell}(k;r,r')$ given in equation (\ref{sesentaidos}).

\begin{eqnarray}\label{setentaisiete}
\frac{X^{2}_{m}u_{m\ell}(k_{m},r)u_{m\ell}(k_{m},r')}
{(E -{\cal E}_{m})^{2}} +
\frac{u_{m\ell}(k_{m},r)\hat{u}_{m\ell}(k_{m},r')+
\hat{u}_{m\ell}(k_{m},r)u_{m\ell}(k_{m},r')}{(E- {\cal E}_{m})}.
\end{eqnarray}
As is evident from the definition (\ref{cuarenta}), the generalized
eigenfunctions $\phi_{n\ell}(k_{m},r)$ and
$\hat{\phi}_{n\ell}(k_{m},r)$ have different dimensions, if one takes
$X_{m}$ of dimension (energy)$^{1/2}$ the normalized eigenfunctions
$u_{n\ell}(k_{n},r)$ and $\hat{u}_{n\ell}(k_{n},r)$ have the same
dimensions namely (energy)$^{-1/2}$ so that when $(X_{m}) =
(energy)^{1/2}$ the higher order Gamow-Jordan vectors become Jordan
vectors with the same dimensions as the Gamow vectors.

This freedom in the normalization rules could be used to define
normalized Gamow-Jordan eigenfunctions with the same dimensions as
those of the Gamow eigenfuntions associated to simple poles of
$G^{(+)}_{\ell}(k;r,r')$.

\section{Summary and conclusions}

In the theory of the scattering of a beam of particles by a short
ranged potential, resonances are asociated to the occurrence of poles
of the scattering matrix $S_{\ell}(k)$, the Green's function
$G^{(+)}_{\ell}(k;r,r')$ and the scattering wave function
$\psi_{n\ell}(k,r)$. These resonance poles are caused by zeroes of the
Jost function lying in the fourth quadrant of the complex $k-$plane.
Accordingly, a degeneracy of resonances, that is, the exact
coincidence of two (or more) simple resonance poles of the scattering
matrix, results from the exact coincidence of two (or more) simple
resonance zeroes of the Jost function, which merge into one double (or
higher rank) zero lying in the fourth quadrant of the complex
$k-$plane.

We found that, associated to a double resonance zero of the Jost
function, there is a Jordan chain of length
two\cite{twentyfive,twentysix,twentyseven} of generalized Gamow-Jordan
eigenfunctions $\{\hat{u}_{m\ell}(k_{m},r), \\ u_{m\ell}(k_{m},r)\}$
belonging to the same degenerate complex energy eigenvalue
${\cal{E}}_{m}$. Hence, the corresponding second rank pole occurring
in the scattering matrix, $S_{\ell}(k)$, the Green's function,
$G^{(+)}_{\ell}(k;r,r')$, and the scattering wave function,
$\psi^{(+)}_{\ell}(k,r)$, is also associated to this Jordan chain of
Gamow-Jordan generalized resonance eigenfunctions.

As the two simple zeroes of the Jost function merge into one double
zero, the two Gamow eigenfunctions corresponding to the two resonances
that become degenerate merge into one Gamow eigenfunction or normal
mode belonging to the double zero of the Jost function. The other
element in the Jordan chain, namely, the Gamow-Jordan generalized
eigenfunction or abnormal mode is not a proper eigenfunction of the
radial Hamiltonian. Hence, at a degeneracy of resonances, one
resonance eigenfunction or normal mode is lost, and a new kind of
generalized resonance eigenfunction or abnormal mode is generated.
Therefore, the dimensionality of the subspace of eigenfunctions
associated to a degeneracy of two resonances or geometric
multiplicity, $\mu_{g}$, of the degeneracy is one, yet, the number of
times the degenerate complex energy eigenvalue is repeated in the
spectral representation of $H^{(\ell)}_{r}$ or algebraic multiplicity
of the degeneracy, $\mu_{a}$, is two. It follows that, the algebraic
multiplicity is larger than the geometric multiplicity of a degeneracy
of resonances.

Explicit expressions for the normalized Gamow and Gamow-Jordan
generalized eigenfunctions in the Jordan chain, written in terms of
the outgoing wave Jost solution, the Jost function and its derivatives
evaluated at the double zero, are obtained from the computation of the
residue of the scattering wave $\psi^{(+)}_{\ell}(k,r)$ function at
the double pole.

We also showed that the Jordan chain of generalized eigenfunctions are
elements of the complex biorthonormal basis formed by the real (bound
states) and complex (resonance states) energy eigenfunctions which can
be completed by means of a continuum of scattering wave functions of
complex wave number. With the help of this result, we derived
expansion theorems (spectral representations) for operators
$f(H^{(\ell)}_{r})$ which are regular functions of the radial
Hamiltonian $H^{(\ell)}_{r}$. In this basis, the operator
$f(H^{(\ell)}_{r})$ is represented by a complex matrix which is
diagonal except for one Jordan block of rank
two\cite{twentyfive,twentysix,twentyseven} associated to the double
zero of the Jost function and the corresponding chain of generalized
eigenvectors. The diagonal entries in this matrix are the eigenvalues
$f({\cal{E}}_{n})$, simple zeroes of the Jost function correspond to
non-degenerate eigenvalues of $f(H^{(\ell)}_{r})$ while the double
zero of the Jost function corresponds to the twice repeated
(degenerate) eigenvalue $f({\cal{E}}_{m})$ in the diagonal entries of
the Jordan block. The off-diagonal, non-vanishing element in this
block is $\partial f({\cal{E}}_{m})/\partial {\cal{E}}_{n}$. In
particular, the occurrence of a double pole in the Green's function,
as function of the complex energy, is thus associated to the
occurrence of a Jordan block of rank two in the complex basis
representation of the resolvent operator and the corresponding Jordan
chain of Gamow-Jordan generalized eigenfunctions.

\section{Acknowledgements}

We thank Prof A. Bohm (U of Texas at Austin) and Prof P. von Brentano
(U zu K\"oln) for many inspiring discussions on this exciting problem.

\end{document}